\documentclass[fleqn,usenatbib]{mnras}
\usepackage[T1]{fontenc}
\usepackage{ae,aecompl}
\usepackage{float}
\usepackage{graphicx}   
\usepackage{color}
\usepackage{amssymb}
\usepackage{amsmath}
\usepackage{enumitem}
\usepackage{booktabs} 

\newcommand{\feh}{$[$Fe/H$]=-1.0$}

\title[Blazhko modulation in the infrared]{Blazhko modulation in the infrared}

\author[J. Jurcsik et al.]{J. Jurcsik$^{1}$\thanks{E-mail: jurcsik@konkoly.hu}, G. Hajdu$^{2,3,4}$, I. D\'ek\'any$^{3}$, J. Nuspl$^{1}$, M. Catelan $^{2,4}$, E. K. Grebel$^{3}$ \\
$^{1}$Konkoly Observatory of the Hungarian Academy of Sciences, H--1525 Budapest PO Box 67, Hungary\\
$^{2}$Instituto de Astrof\'isica, Pontificia Universidad Cat\'olica de Chile, Av. Vicu\~na Mackenna 4860, 782-0436 Macul, Santiago, Chile\\
$^{3}$Astronomisches Rechen-Institut, Zentrum f\"ur Astronomie der Universit\"at Heidelberg, M\"onchofstrasse 12-14, 69120\\ Heidelberg, Germany\\
$^{4}$Millennium Institute of Astrophysics, Santiago, Chile\\
}

\date{Accepted 2018 January 10, Received 2017 December 9, in original form }

\pubyear{2018}

\begin{document}
\label{firstpage}
\pagerange{\pageref{firstpage}--\pageref{lastpage}}
\maketitle

\begin{abstract}
 We present first direct evidence of modulation in the $K$-band of Blazhko-type RR Lyrae stars that are identified by their secular modulations in the I-band data of OGLE-IV.  A method has been developed  to decompose the $K$-band light variation into two parts originating  from the temperature and the radius changes using synthetic data of atmosphere-model grids. The  amplitudes of the temperature and the radius variations derived from the method for non-Blazhko RRab stars are in very good agreement with the results of the Baade-Wesselink analysis of RRab stars in the M3 globular cluster confirming the applicability and correctness of the method. 
It has been found that the Blazhko modulation is primarily driven by the change in the  temperature variation. The radius variation plays a marginal part, moreover it has an opposite sign as if the Blazhko effect was caused by the radii variations.
This result reinforces the previous finding based on the Baade-Wesselink analysis of M3 (NGC~5272) RR Lyrae, that significant modulation of the radius variations can only be detected in radial-velocity measurements, which relies on spectral lines that form in the uppermost atmospheric layers. Our result gives the first insight into the energetics and dynamics of the Blazhko phenomenon, hence it puts strong constraints on its possible physical  explanations.

\end{abstract}

\section{Introduction}\label{intro.sec}
The Blazhko effect, the periodic/cyclic amplitude and phase changes of the pulsation light curve of RR Lyrae (RRL) stars on time scales of tens to thousands of days  \citep{blazhko,shapley}, is a  century-old puzzle of stellar pulsation. The properties of the modulation have already been documented in detail, based on extended photometric time-series observations of large samples of Blazhko stars. The ground-based observations of the Konkoly Blazhko Survey \citep{kbs,kbsaip}, from  Antarctica  \citep{c14},  the Optical Gravitational Lensing Experiment (OGLE) project \citep{smolec15,ps17}, and the $CoRoT$- and $Kepler$-satellite measurements \citep{szkm,k11,kepler,molnarpl} have revealed several new features of the modulation. Despite these achievements, fundamental understanding of the phenomenon is still missing. Summaries of our recent knowledge on the Blazhko phenomenon and reviews on the problems in the interpretations  were published  in  recent years by  \cite{sz14,kgrev,smolecbl}, and \cite{kollathpl}.

Information on the modulation properties of RRL stars in the near-infrared is very sparse.
\cite{sol08} measured time-series data of RR Lyr itself in the $JHK$ bands, and they attributed the observed increased scatter of the light curves to the Blazhko effect. However, looking at the light curves \citep[figure 1 in][]{sol08} it can be seen that despite the  similarity of the amplitudes in the three bands, the $K$-band light curve shows the smallest scatter. Therefore, if modulation is the reason of the scatter than it has the most-reduced amplitude in the $K$-band.

The $K_\mathrm{S}$-band (hereafter $K$) observations of RRL stars in $\omega$ Centauri \citep[NGC~5139][]{n15} have not firmly established $K$-band modulation, either. The limited amount of data points, and the large observational base-line hampered these efforts, although an increased scatter was also noticed by \cite{n15} for some of the $\omega$ Cen variables.

Spectroscopic observations of Blazhko stars  \citep[e.g.][]{psp,cc06,cvg08,cp13,m3data} show that their  radial-velocity variation  is also modulated and displays parallel changes   with  the  amplitude and phase modulations of the light-curve. However, the first attempt to perform a Baade-Wesselink analysis of Blazhko stars has led to the surprising conclusion that the radial displacement of the photosphere does not seem to show any variation even for stars with large modulation amplitude \citep{m3bw}. 

The possible lack of the modulation in the $K$ band was interpreted as a depth-dependent property of the Blazhko effect by \cite{m3bw}. While the radial-velocity data document the dynamical movement of the higher, line-forming layers of the atmosphere, the $K$-band light curve reflects the changes of the deepest regions of the photosphere due to the near-infrared dip in the opacity. 

However, because questions remain about $K$-band modulation amplitudes of Blazhko stars, an independent investigation of Blazhko modulation in the $
K$-band is desirable.

 The OGLE-IV observations \citep{ogleIV} provide excellent time-series data of a huge number of  RRL stars in the Galactic bulge, and $\sim40$ per cent of these stars show the Blazhko effect \citep{ps17}. Selecting a sample of Blazhko RRL stars from the OGLE-IV data base, our aim was to check the $K$-band light curves of the  VISTA Variables in the V\'ia L\'actea (VVV, \citealt{vvv}) survey. Based on the results of our analysis, the properties of the modulation in $K$ band have been explored.

\section{Data}\label{data.sec}

\begin{figure}
\centering
\includegraphics[width=8.5cm]{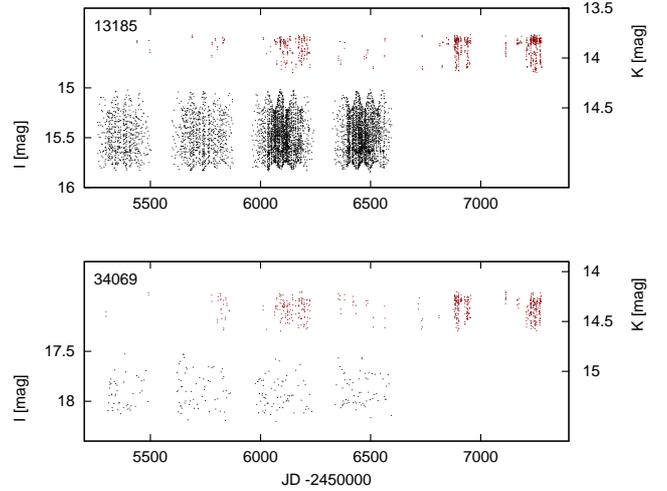}
\caption{The OGLE $I$-band and the VVV $K$-band measurements of two Blazhko stars in the sample are shown to document the time-span and coverage of the data. ID\,13185 and  ID\,34069 have  the largest and the smallest number of observations in the sample of stars used for the detailed analysis.}
\label{time} 
\end{figure}

\begin{figure}
\centering
\includegraphics[width=9.2cm]{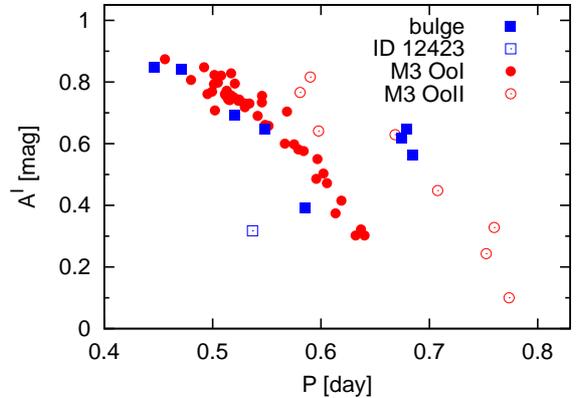}
\caption{Peak-to-peak $I$-band amplitudes versus pulsation period are shown for the stable-light-curve, reference sample of stars. For comparison, the $A^I$ versus $P$ relation  of non-Blazhko RRab stars in the M3 globular cluster is also plotted.}
\label{iamp} 
\end{figure}

\begin{table} 
\begin{center} 
\caption{List of the stars selected for the detailed analysis\label{data}} 
\begin{tabular}{lcrc@{\hspace{1mm}}c}
\hline  
Var.&$P_{\mathrm{puls}}$ & $P_{\mathrm{mod}}$ & $E(I-K)$$^*$ & ${R_0}^{**}$ \\ 
OGLE ID& [d]&  [d]\,\,& [mag] & $\mathrm{R_{\odot}}$ \\ 
\hline
\multicolumn{5}{l}{Stable RRab stars}\\
10879  & 0.470743  && 1.195  & 4.54\\
11025  & 0.684346  && 1.105  & 5.51\\
11105  & 0.520411  && 0.780  & 4.78\\
12423  & 0.536823  && 0.971  & 4.86\\
13714  & 0.548255  && 1.394  & 4.91\\
13851  & 0.674377  && 1.088  & 5.47\\
14408  & 0.446024  && 0.912  & 4.41\\
33548  & 0.585278  && 2.202  & 5.08\\
34927  & 0.679371  && 1.413  & 5.49\\
\multicolumn{5}{l}{Blazhko RRab stars}\\
08263  & 0.632747&	35.82  & 2.155  & 5.29\\
09904  & 0.490706&	111.35 & 1.283  & 4.64\\
10402  & 0.552169&	91.08  & 1.522  & 4.93\\
11000  & 0.586359&	56.81  & 0.869  & 5.09\\
11104  & 0.540359&	185.09 & 1.105  & 4.88\\
11134  & 0.656004&	177.15 & 1.133  & 5.39\\
11381  & 0.583529&	107.84 & 0.935  & 5.07\\
11992  & 0.613531&	73.44  & 0.746  & 5.21\\
12085  & 0.547507&	28.56  & 0.774  & 4.91\\
12088  & 0.557407&	31.26  & 0.741  & 4.95\\
13185  & 0.620459&  	57.87  & 0.812  & 5.24\\
13640  & 0.588433&      476.19 & 0.745  & 5.10\\
14225  & 0.562246&	193.46 & 0.864  & 4.98\\
14261  & 0.510530&	201.6  & 1.017  & 4.73\\
14322  & 0.581213&	26.54  & 1.053  & 5.06\\
14350  & 0.479727&	37.05  & 1.089  & 4.58\\
14654  & 0.540359&	18.96  & 1.736  & 4.87\\
14859  & 0.447746&	68.97  & 0.993  & 4.42\\
16012  & 0.517044&      64.93  & 1.180  & 4.76\\
34069  & 0.569616&	135.32 & 2.820  & 5.01\\
34864  & 0.559212&	26.04  & 1.451  & 4.96\\
35076  & 0.491096&	108.44 & 1.606  & 4.64\\
\hline
\multicolumn{5}{l}{$^*$ $E(I-K)$ is derived from the calibration of the $M_I$ and $M_K$}\\ 
\multicolumn{5}{l}{absolute magnitudes of RRL stars \citep{catelan}}\\
\multicolumn{5}{l}{transformed to the photometric system of VVV.}\\
\multicolumn{5}{l}{$^{**}$ The radii are calculated from the log$R/R_{\odot}$($\log p$, $\log Z$)}\\
\multicolumn{5}{l}{calibration of \citet{marconi}.}
\end{tabular} 
\end{center}
\end{table}

\begin{figure*}
\centering
\includegraphics[width=18.8cm]{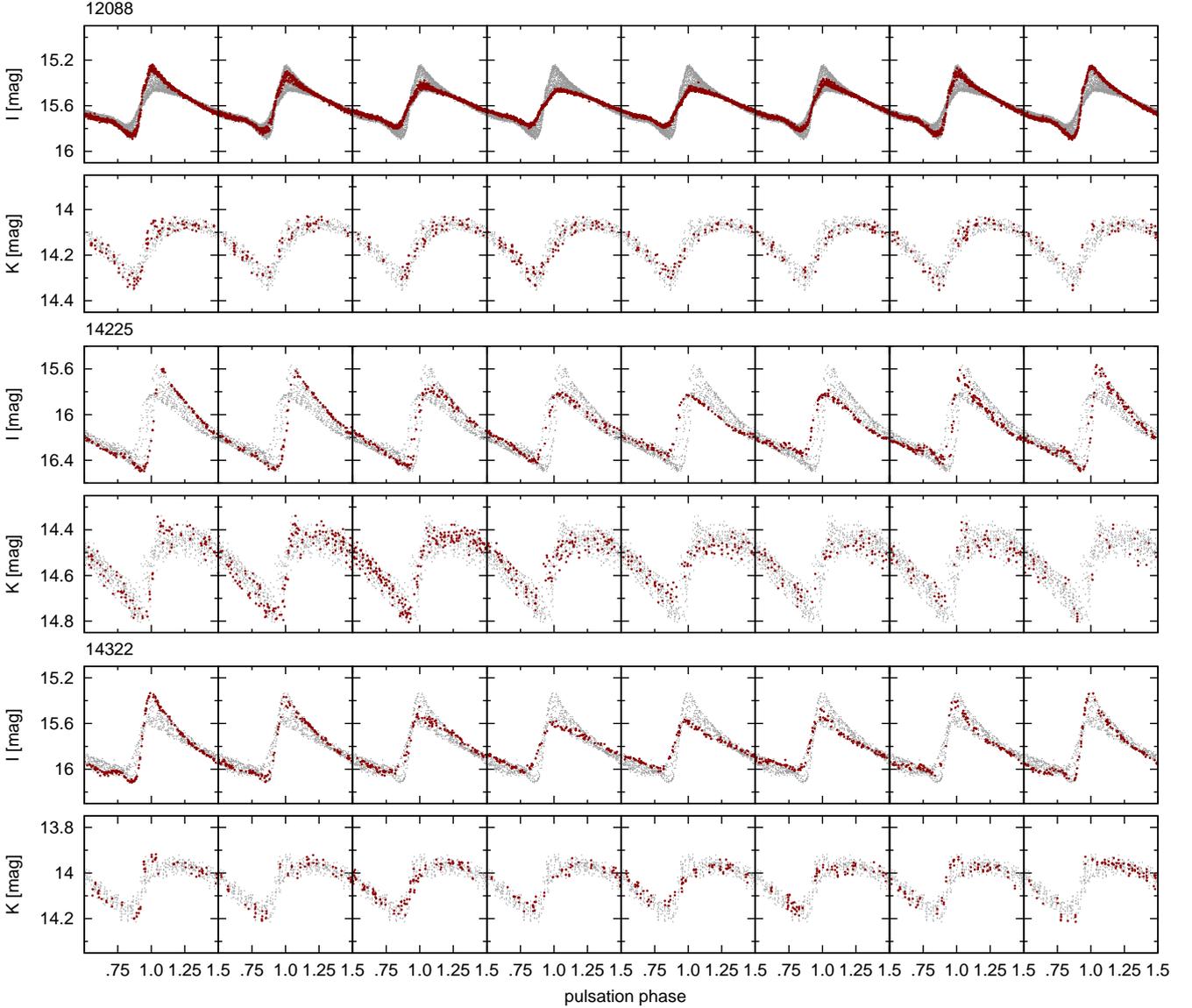}
\caption{OGLE $I$-band and VVV $K$-band light curves of three Galactic-bulge RRab stars in eight phases of the Blazhko modulation. The full light curves are shown   in grey in the frames for comparison. Large-amplitude modulation is evident in the $I$ band and parallel changes  appear also in the $K$ band at around the maximum phase of the pulsation.  Phase 1.0 (0.0) corresponds to the maximum phase of the mean light curve in the $I$ band.} 
\label{lck} 
\end{figure*}

The OGLE-IV $I$-band photometry \citep{ogleIV} has been utilised in conjunction with the
VVV $K$-band observations of the Galactic bulge in the analysis. 

We have used photometry based on individual detector frame stacks (pawprints) from the VVV survey, provided by the VISTA Data Flow System \citep{emerson, irwin} of the Cambridge Astronomy Survey Unit (CASU). Since the detection of light-curve modulation  is generally very sensitive to the number of data points, we have focused on Blazhko variables, selected from the OGLE-IV RR~Lyrae data base\footnote{\url{http://ogle/astrouw.edu.pl/}}, which had more than the usual number of data points ($50-70$) in the VVV survey.  At first, we selected an initial sample of 46  Blazhko variables with a sufficient number of  $K$-band data ($>500$) to detect any possible modulation of the light curves.

As CASU provides photometry with multiple aperture sizes, the optimal apertures were chosen based on the  scatter of the observations.  The light curve with the aperture yielding the smallest scatter was selected individually for the studied stars. 

VISTA observations of a single tile (a contiguous area of $\sim 1.5^{\circ} \times 1^{\circ}$) constitute 6 separate sub-exposures called pawprints, in order to fill the substantial gaps between the detectors of VIRCAM. These data clusters span only ~3-min intervals. Furthermore, neighbouring tiles have overlapping regions near the edges. Data points belonging to the same
pawprint (and tile) observations were identified and corrected for  any possible small offsets arising from systematic differences in the photometric calibration in the order of $0.005-0.03$. mag.

The partially overlapping OGLE-IV  and  VVV observations  cover the JD $2\,455\,300-2\,457\,280$ and JD $2\,455\,260-2\,456\,600$ time intervals, respectively (see Fig.~\ref{time}). Although this makes  a combined analysis of the OGLE-IV and VVV data feasible, the phased $I$- and $K$-band light-curves  do not match in phase accurately enough for many variables, most probably due to the strong changes in the pulsation period of these stars. To successfully carry out the analysis, it is required for the variable in question to possess stable pulsation and relatively simple modulation behaviour, and the data to be extensive enough  to allow accurate light curve decomposition. 
Therefore, stars with a large period-change rate and/or with strongly multi-periodic modulation have been omitted from the sample. Some stars with inadequate $K$-band data  were removed, too. 
This selection made it possible to perform the analysis 
using the same pulsation and modulation frequencies in both bands.

After evaluating each variable by these criteria, we were left  with  22 Blazhko stars for the detailed analysis.
The time coverage of the observations for the stars with the largest (OGLE-BLG-RRLYR-13185, hereinafter OGLE-BLG-RRLYR is shortened to ID) and the smallest (ID\,34069) number of data points in this final sample is documented in Fig.~\ref{time}.

For reference, nine RRab stars with stable light curves (i.e. non-Blazhko) have also been included in the  study, in order to check and validate the results of the analysis.
The data of RRab stars in the M3 globular cluster \citep{m3data,m3bw} are used in the course of the analysis in many contexts. Although the metallicity of M3 is ${\rm [Fe/H]} = - 1.5$, i.e. smaller than the mean metallicity of the bulge RRL stars  \citep[\feh,][]{piet15}, multi-colour information on the modulation properties and radius- and temperature-change amplitudes  have only been published for a large, homogeneous sample of RRab stars in \cite{m3data}.

 The $I$-band  amplitude versus period relation  of this reference sample of bulge stars is compared to the results of stable-light-curve RRab stars in M3 in Fig.~\ref{iamp}.   The location of the bulge stars in this plot indicates that they are similar to the variables in M3, i.e. the sample contains both Oosterhoff I- and II-type stars \citep[OoI and OoII, respectively;][]{oo}. The OoII stars have larger amplitudes than the OoI-type stars at the same period \citep[for a revirew of the Oosterhoff effect read e.g.,][]{csbook}. This is in line with the detection of OoI and OoII populations in the bulge by \cite{prudilpl}. An ``outlier" (ID\,12423) can also be seen in this plot. Based on its position, this star has either a significantly larger metallicity than the others or its photometry is erroneous. However, checking the finding chart of ID\,12423 in the OGLE data base, we have not seen any indication for any defect of the photometry. Thus ID\,12423 is regarded as an example for the inhomogeneity of the bulge sample, or this star does not belong to the bulge population.

Table~\ref{data} lists the OGLE IDs of the variables and their pulsation and modulation periods determined and used in the analysis. The $E(I-K)$ reddenings are derived using the $M_I$ and $M_K$ absolute magnitudes of RRL stars \citep{catelan} transformed to the VISTA photometric system, and are given in column 4 of Table~\ref{data}. The radius values ($R_0$, column 5) are calculated according  to the formula: $\log R/R_\odot=0.749(\pm0.006)+0.52(\pm0.03)\log P-0.039(\pm0.006)\log Z$ \cite[Eq. 7 of][]{marconi}, adopting $Z=0.001$.

\section{Light-curve analysis}\label{lc.sec}

The light curves were analysed using  the program packages MUFRAN \citep{
mufran} and LCfit \citep{nl}. First, the solution of the $I$ data was determined as the amplitudes are larger and the noise is smaller in the $I$ band than in $K$. Using  sine decomposition, the light curves are described as Fourier series of the pulsation  ($kf_p, k=1...i$) and modulation ($kf_p\pm lf_m,\,\, k=0....j, l=1...4$) frequency components and the mean magnitude ($a_0$). The actual pulsation/modulation-component  content  has been determined separately for each star; only frequency components with amplitudes larger than $3 \sigma$ are taken into account.

After determining the period of the modulation for the $I$-band data, a minor but clear sign of changes of the light-curve shape was detected in the $K$-band at different phases of the modulation. The modulation is evident not only for the 22 stars selected for the analysis but for all of the 46 Blazhko stars for which the $K$-band light curves were evaluated. 

Fig.~\ref{lck} documents the $I$- and $K$-band light curves of three representative Blazhko stars of the sample in eight phases of the modulation cycle. The complete, phased light curves are  shown in grey in each panel of Fig.~\ref{lck} for comparison.  Although the modulation has only a small amplitude in the $K$ band, there is no doubt that the Blazhko modulation appears in this band, too. In view of this, it seems that the absence of clear-cut modulation in the $K$-band data of \cite{sol08} and \cite{n15} was likely due to the sparseness and small S/N of their data, along with the intrinsically small amplitude of the Blazhko signal in the near-IR.

The $f_p$  and $f_m$ frequencies  determined for the $I$-band data were accepted to be valid for the $K$-band light-curve solution.  However, it contains a smaller number of the pulsation and modulation components than in the $I$ band because of the  more sinusoidal shape of the light curve and of the small amplitude of the modulation in this band.

Non-Blazhko RRab stars were analysed similarly; the adopted pulsation frequency was determined for the $I$-band data.

Once the light-curve solutions of the data sets were determined,  simultaneous $I$- and $K$-band time-series data were generated, and the synthetic $K$ light and the $I-K$ colour curves were used in the course of the analysis.

\subsection{Amplitude ratios of the $I$- and $K$-band light curves}\label{amp.sec}

\begin{table} 
\begin{center} 
\caption{Mean values of the $A^K/A^I$ amplitude ratios of the pulsation and modulation frequency components of Blazhko and non-Blazhko samples of stars in the  bulge and for the $A^I/A^V$ ratios in M3.\label{ak}} 
\begin{tabular}{l@{\hspace{3mm}}c@{\hspace{3mm}}l@{\hspace{1mm}}cl}
\hline  
Order& Freq.&Star& $\overline {A^K/A^I}$(rms)$^{*}$&\multicolumn{1}{c}{$\overline {A^I/A^V}$}(rms)\\
&&&Galactic bulge &\multicolumn{1}{c}{M3}\\ 
\hline
1st&&&&\\
&$f_p$ &stable& 0.58(3) & 0.635 (2)\\
&$f_p$ &Bl   & 0.57(2) & 0.623 (2)\\
&$f_m$ &Bl    & 0.30(2) & 0.630(10)\\
2nd&&&&\\
&$f_p$ &stable& 0.41(2) & 0.653 (2)\\
&$f_p$ &Bl     & 0.40(1) & 0.633 (4)\\
&$f_m$ &Bl     & 0.32(2) & 0.625(15)\\
3rd&&&&\\
&$f_p$ &stable& 0.39(2) & 0.665 (3)\\
&$f_p$ &Bl     & 0.37(1) & 0.649 (7)\\
&$f_m$ &Bl     & 0.33(2) & 0.661(14)\\
\hline
\multicolumn{5}{l}{$^*$ The rms scatter corresponds for the last digit(s).}\\ 
\end{tabular} 
\end{center}
\end{table}

\begin{figure}
\centering
\includegraphics[width=8.7cm]{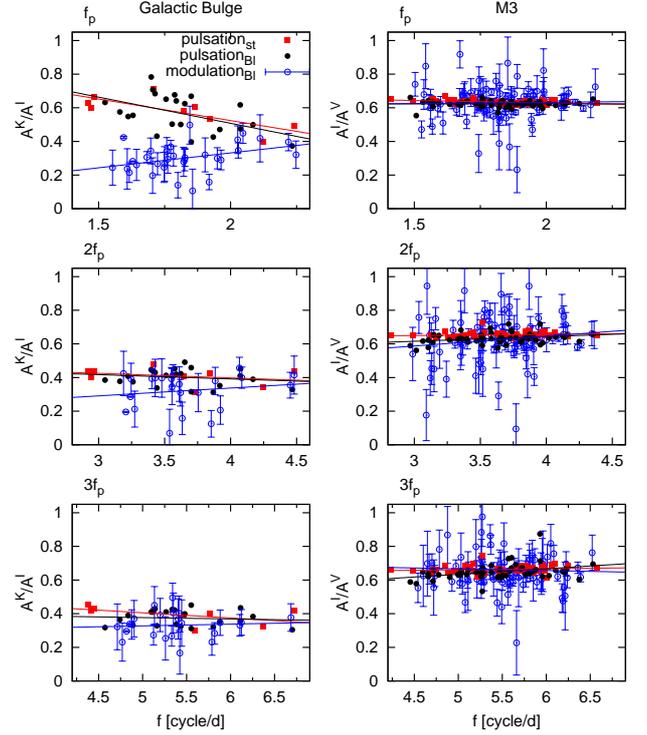}
\caption{Amplitude ratios of the  $K$ and $I$ band ($A^K/A^I$) pulsation ($kf_p$, filled symbols) and modulation ($kf_p\pm f_m$, open symbols) frequency components  of the Galactic bulge RRab stars are plotted in the left-hand panels. The amplitude ratios of the pulsation component of stable and Blazhko RRab stars are shown by different symbols. The first three harmonic-order components are shown in the top, middle, and bottom panels separately.
 The right-hand panels show the same amplitude ratios for the $V$- and $I$-band data of variables in the M3 globular cluster.
The errors of the amplitude ratios of the pulsation components are not shown as they hardly exceed the points' size.
Linear fits of the data of each sample are drawn in the plots.  }
\label{foura} 
\end{figure}

\begin{figure}
\centering
\includegraphics[width=8.7cm]{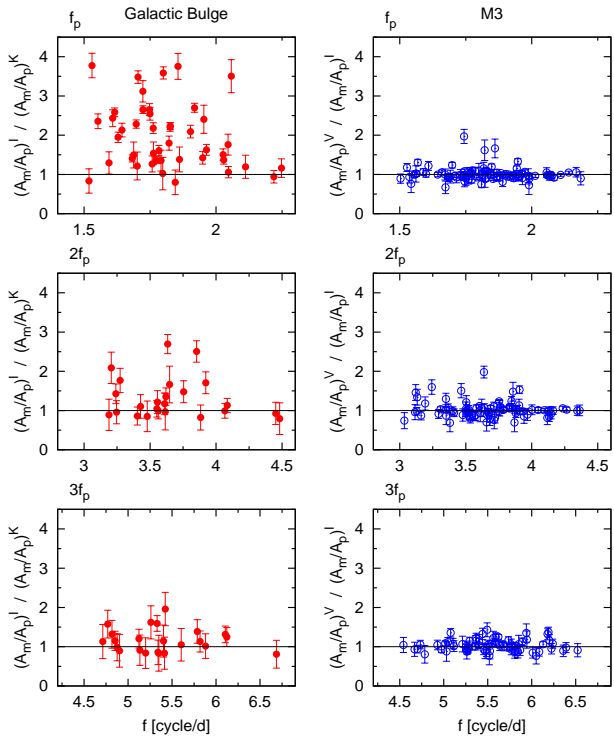}
\caption{The ratio of the $I$- and $K$-band relative strength of the modulation ($A_m/A_p$) of Blazhko stars in the Galactic bulge is shown for the first three harmonic orders  in the left-hand panels. The right-hand plots display the same ratios for the $V$ and $I$ data of the M3 Blazhko stars. For guidance, the equality of the values of the relative strength of the modulation in the two bands are indicated by horizontal lines in each panel.}
\label{fouram} 
\end{figure}

The light curves of RRL stars are significantly different in the $I$ and $K$ bands. The optical $BVRI$ light curves of RRab stars have similar, strongly asymmetric  shapes with a sharp rising branch and a peaked maximum (with the exception of the longest-period, smallest-amplitude variables), in contrast with  the  somewhat less asymmetric light curves with flat maximum seen in the $K$-band (see Fig.~\ref{lck}). Besides the light-curves' shapes, the characteristics of the modulation seem to be different in the $I$ and $K$ bands, too.   The modulation affects nearly each phases of the pulsation with the most significant changes appearing at maximum and minimum phases in the $I$ band. On the contrary, profound changes of the $K$-band light curves are evident only around the maximum.

The differences between the  pulsation and modulation properties in the $K$ and $I$ bands are investigated by checking the differences between the amplitude-ratio relations of pulsation and modulation frequency components in different orders  in the two bands.
The results are compared to the amplitude ratios for the   $I$- and $V$-band light curves  of  RRab stars of the M3 globular cluster published in  \cite{m3data}. In order to estimate the significance of these relations, the errors of the amplitude ratios are calculated using the formal errors of the Fourier amplitudes. We note, however, that these formal errors may underestimate the true errors as they do not reflect any uncertainty originating from any systematic defect of the photometry.

As the pulsation amplitude in the $K$ band is about half of the amplitude in the $I$ band, we checked first whether the same rate of  decline in the amplitudes of the modulation frequency components are observed.
Hence, the changes of the $A^K/A^I$ amplitude ratios of the pulsation ($p$) components of non-Blazhko and Blazhko stars are compared with the amplitude ratios of the modulation ($m$) frequency components. The results for the first three harmonic orders are  shown in the left-hand panels of Fig.~\ref{foura}. The right-hand panels display similar amplitude ratios for the M3 data, which relate the decline of the pulsation and modulation amplitudes between the $V$ and the $I$ bands.

The $A^K/A^I$ amplitude ratios determined for the bulge sample show significant differences, especially in the first harmonic order. Although the $A^K/A^I$  amplitude ratios  of the pulsation components are the same for the stable and for the Blazhko type stars (shown by filled squares and circles, respectively, in Fig.~\ref{foura}) and the linear fits to these data are in coincidence in each harmonic order, they differ significantly  from the $A^K/A^I$  amplitude ratios of the modulation components (shown by empty circles in Fig.~\ref{foura}) as can be seen in the plots. The amplitude ratios of the pulsation and modulation components show similar variations in the second/third harmonic orders as in the first harmonic order, but  the differences become less and less significant.

The mean values of the different amplitude ratios of the bulge and the M3 samples are listed in Table~\ref{ak}. The differences between the mean values of the  $A^K/A^I$ amplitude ratios of the pulsation and the modulation components in the bulge sample are 0.27, 0.08, and 0.05 in the first-, second-, and third-harmonic orders, respectively, but there is no difference between the similar ratios for the $I$- and $V$-band M3 data. 

The small value of the $A^K/A^I$ ratios of the modulation components compared to the amplitude ratios of the pulsation components proves that the decrease of the modulation amplitudes between the $I$ and $K$ bands is indeed more significant than the decrease of the amplitudes of the pulsation components. As this is most prominent in the first harmonic order,  the modulation of the sinusoidal-shaped part of the light curve is the most anomalously reduced in the $K$ band.

Another interesting feature of the  $A^K/A^I$ ratios of the pulsation and the modulation frequencies is that they show  opposite-sign period dependence. 
The first order $A^K/A^I$  ratio of the pulsation components is  as large as 0.6 at small pulsation frequencies (long periods) but it is around 0.4 at high frequencies (short periods). Meanwhile, the amplitude ratio of the modulation component  increases from 0.2 to 0.4. The largest discrepancy between the $A^K/A^I$ ratio of the pulsation and the modulation components is thus detected at the short-frequency (long-period) end.

Secondly, we examine the relative strength of the modulation to the pulsation in the different bands.
The relative strength of the modulation is measured  by the $A_m/A_p$ ratio of the  modulation  and the pulsation frequency components. 
To follow  its changes in the different bands the $(A_m/A_p)^I/(A_m/A_p)^K$ ratios determined for the bulge sample are compared to the similar ratios for the $V$ and $I$ bands of the M3 data in Fig.~\ref{fouram}. 

The M3 data indicate that these ratios are the same in the $V$ and the $I$ bands; the mean values of the $(A_m/A_p)^V / (A_m/A_p)^I$ ratios are  1.0 within the error range in each of the first three harmonic orders (right-hand panels of Fig.~\ref{fouram}). The same ratios for the $I$- and $K$-band amplitudes (left-hand panels in Fig.~\ref{fouram}) are, however, significantly larger than 1.0, indicating that the relative strength of the modulation to the pulsation can be even about $3-4$ times larger in the $I$ band  than in the $K$ band.
The difference between the relative strength of the modulation in the $I$ and $K$ bands is again the largest in the first harmonic order.

Summarising, the amplitude-ratio relations of the pulsation- and modulation-frequency components show different characteristics   in the $K$ band compared to the properties in the $V$ and $I$ bands. 
The $K$ amplitudes of the modulation components do not follow the same colour dependence as the pulsation, their amplitudes are smaller than expected, as indicated by the low values of their $A^K/A^I$ ratios, and the large values of the $(A_m/A_p)^I / (A_m/A_p)^K$ ratios. Although there is no doubt that the modulation is also present in the $K$ band, these results show that the amplitudes of the modulation frequency components are anomalously reduced.
The most discrepant behaviour of the modulation components is observed in the first harmonic order.

\section{Separation of the $K$-band light curve into  temperature- and radius-change  connected variations}\label{method}

\begin{figure*}
\centering
\includegraphics[width=17.7cm]{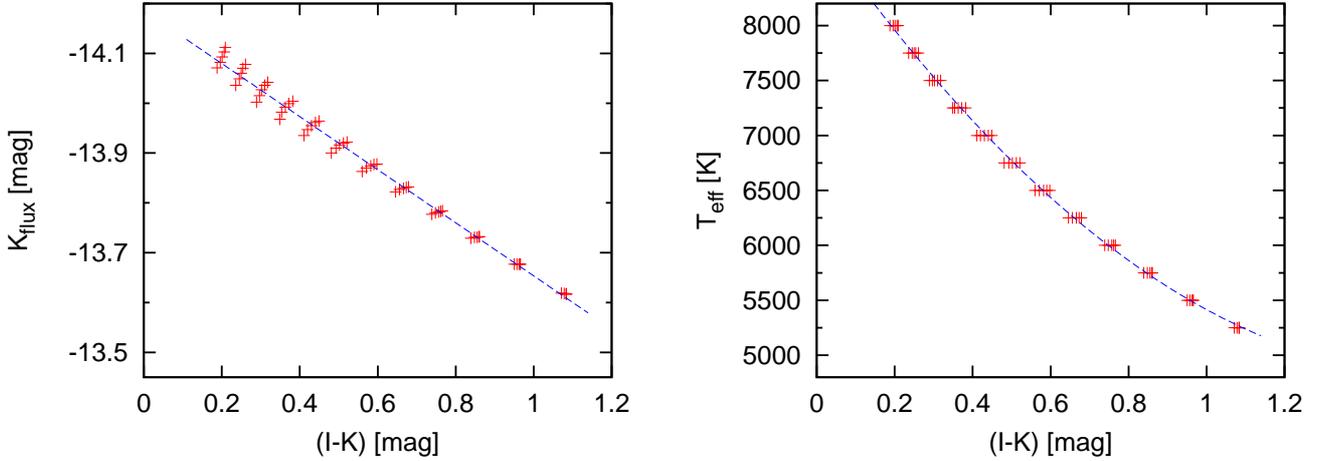}
\caption{The linear fit of the $K_{\mathrm{flux}}$  and quadratic fit of the temperature  to the $I-K$ colour index  are shown in the left and right  panels, respectively. Synthetic data of $\alpha$-enhanced, $\mathrm{[M/H]}=-1$ atmosphere models \citep{kurucz} in the $T = 5250-8000$, $\log g=2.0-4.0$ ranges are used.}
\label{kurucz} 
\end{figure*}

\begin{figure*}
\centering
\includegraphics[width=17.7cm]{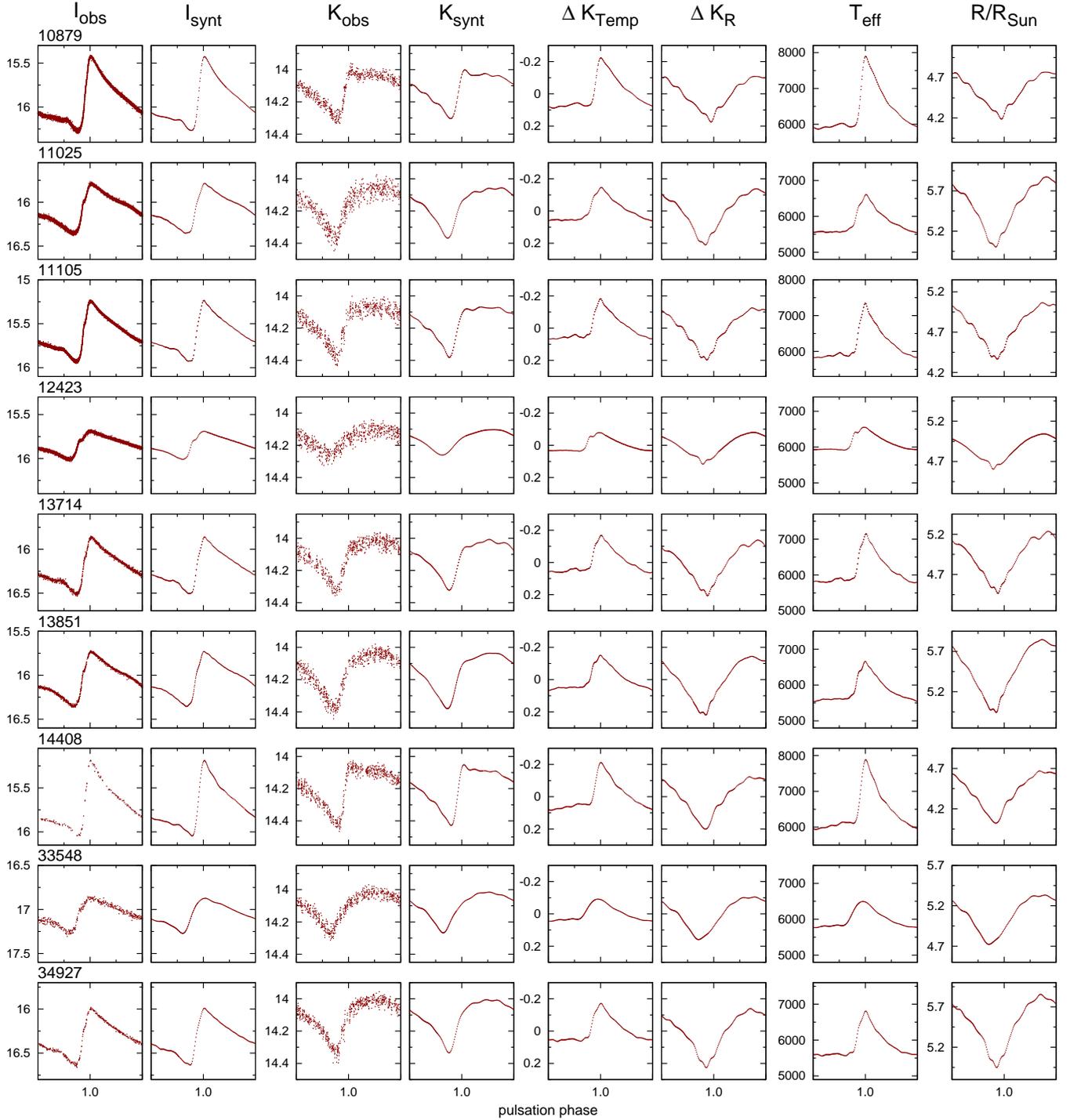}
\caption{Results for non-Blazhko RRab stars of the Galactic-bulge sample are documented in the figure.  Observed and synthetic $I$- and $K$-band phased light curves are shown in the four left-hand columns. The fifth and sixth columns document the decomposed $\Delta K$ variations connected to the temperature and to the radius changes. The estimated temperature and radius changes are shown in the last two columns.}
\label{lcst} 
\end{figure*}

\begin{figure*}
\centering
\includegraphics[width=17.6cm]{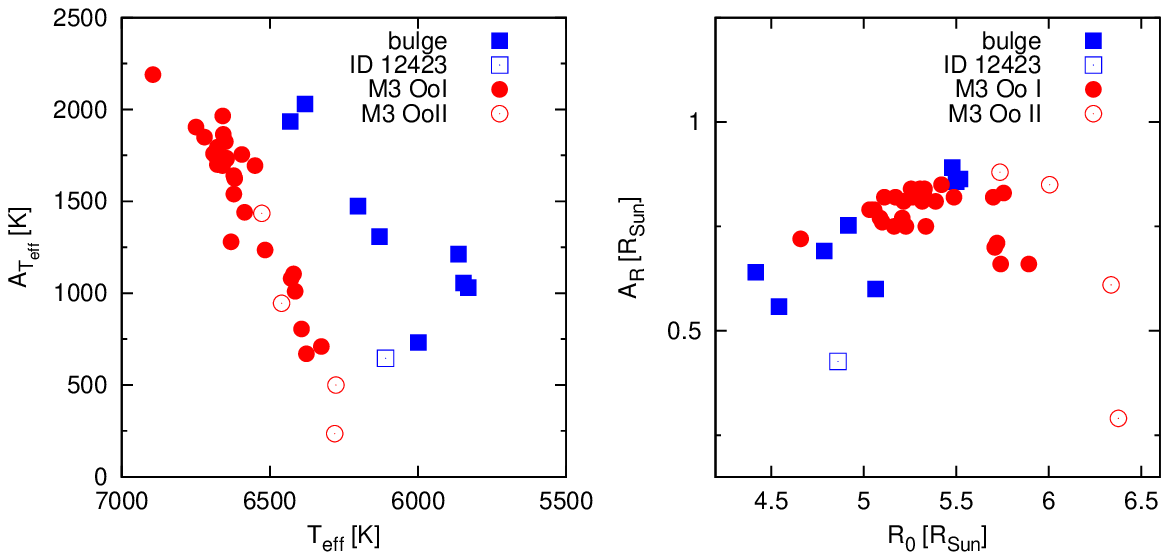}
\caption{Comparison of the derived amplitudes of the temperature and radius changes of stable-light-curve Galactic-bulge RRL stars with the results of the Baade-Wesselink analysis of variables in the M3 globular cluster \citep{m3data}. The $A_{T_{\mathrm{eff}}}$ and $A_{R/R_{\odot}}$ amplitudes correspond to the total amplitudes of the variations. }\label{comp} 
\end{figure*}

Our main goal   is to decompose the $K$-band light curve into parts originating from the change in radius ($\Delta K_R$) and the change in temperature ($\Delta K_{T_\mathrm{eff}}$)  using the $I$ and $K$ observations and combining them with the synthetic data of model-atmosphere grids \citep{kurucz}. 

The monochromatic flux received at the Earth ($f_{\lambda}$) depends on the angular radius ($\theta=R/d$, where  $R$ and $d$ are the radius and the distance of the star) and the monochromatic  flux at the stellar surface ($F_{\lambda}$):
\begin{equation}\label{eq:base}
{f_{\lambda}={\theta}^2 F_{\lambda}}. 
\end{equation} 
Integrating this relation for a given photometric band using  appropriate filter functions, the observed dereddened fluxes can be related to the stellar flux directly.

As we are primarily interested  in the relative changes of the parameters and not in their absolute values, we use the following, pulsation phase-dependent  form of Eq.~\ref{eq:base}:
\begin{equation}\label{eq:flux}
{f_K(\varphi)\over{\overline{f_K}}}=\left[{\theta(\varphi)\over{\overline{\theta}}}\right]^2 {{{F_K}^{\mathrm{model}}}(\varphi)\over {\overline{{f_K}^{\mathrm {model}}}}}, 
\end{equation} 
where $f_K$ is the observed, dereddened $K$-band integrated flux, and ${F_K}^{\mathrm{model}}$ is the corresponding value computed from stellar atmosphere models with appropriate physical parameters ($T_{\mathrm {eff}}$, $\log g$, [M/H]); $\varphi$ denotes the pulsation phase and the bar indicates the mean value.

Transforming Eq.~\ref{eq:flux} to magnitude scale, 
\begin{equation}\label{eq:mag}
{\Delta K(\varphi)= -5 \log \left[R(\varphi)/{R_0}\right]+\Delta {K_{\mathrm{model}}}(\varphi)}
\end{equation} 
is obtained, where $\Delta$ denotes the difference between the actual value at pulsation phase $\varphi$ and the mean; and $R_0$ is the mean radius of the star.

The advantage of using $K$-band magnitudes is that $\Delta {K_{\mathrm{model}}}$ is a linear function of the colours with good approximation, and it depends only marginally on the  $\log g$ in the relevant parameter range ($T_{\mathrm {eff}} = 5250-8000$, $\log g=2.0-4.0$, $\mathrm{[M/H]}=-1$, $\alpha$-enhanced). The  effect of $\log g$  on the $K_{\mathrm{model}}$ values varies from 0.0 to 0.06 mag in the $(I-K)=1.0-0.2$ mag colour index interval. 
This means that the disregard of the $\log g$ dependence influences the results by more than 0.02 mag only at temperatures hotter than $\sim6750$ K, i.e. at around brightness maximum. 

The linearity of the relation between the $K_{\mathrm{model}}$ magnitude and the $I-K$ colour index  for the parameter range considered using  the combined $coubes$ and $rijkl$ FRIDM10AODFNEW model grids of \cite{kurucz} is documented in the left-hand panel of Fig.~\ref{kurucz}. This relation is 
\begin{equation}\label{eq:kik}
 K = 0.533(I-K)-14.186,
\end{equation} 
with rms$=0.015$ mag. We also note that the linearity of Eq.~\ref{eq:kik}  guarantees that the derived $\Delta K_{\mathrm{model}}(\varphi)$ does not depend on the uncertainties of the used  $E(I-K)$ reddening values or on the mean values of the observations.

As a consequence of neglecting the $\log g$ dependence of the $K_{\mathrm{model}}$ values, the two terms in the right-hand side of Eq.~\ref{eq:mag} can be regarded as  the radius-change ($\Delta K_R$) and the temperature-change ($\Delta K_{T_{\mathrm {eff}}}$) induced parts of the observed light variation with a good approximation, i.e. $\Delta K_{T_{\mathrm {eff}}} = \Delta K_{\mathrm{model}}$.

The implemented procedure consists of the following steps:

\begin{enumerate}[label=(\roman*)]
\item{  The $\Delta K_{T_{\mathrm {eff}}}(\varphi)$ values are determined using Eq.~\ref{eq:kik} and the observed $\Delta [I-K](\varphi)$ dereddened magnitudes, obtained from the synthetic $I(\varphi)$ and $K(\varphi)$ light curves and the $E(I-K)$ reddening values listed in Table~\ref{data}.}
\item{ Then, $\Delta K_R(\varphi)$ is obtained from:
\begin{equation}\label{eq:r}
\Delta K_R(\varphi)=\Delta K_{\mathrm {obs}}(\varphi)- \Delta K_{T_{\mathrm {eff}}}(\varphi).
\end{equation}}
\item{Finally, the temperature and radius changes are estimated on an absolute scale. 
$T_{\mathrm {eff}}(\varphi)$  is derived from the colour-temperature relation given by the model-atmosphere grid as shown in the right-hand panel of Fig.~\ref{kurucz}. It has the form of:
\begin{equation}
 T_{\mathrm{eff}}=1568(I-K)^2-5059(I-K)+8907,
\end{equation} 
with rms$=42$ K. Because this relation is nonlinear, the derived  $ T_{\mathrm {eff}}(\varphi)$ values are biased by the uncertainty of the $E(I-K)$ reddening values. 

To transform the $\Delta K_R(\varphi)$ to the radius change, we need to know the mean radius values ($R_0$) of the stars,  as 
\begin{equation}\label{eq:dkr}
\Delta K_R(\varphi)= -5 \log(R(\varphi)/R_0),\,\, \mathrm{i.e.} \,\,
 R(\varphi)=R_0 10^{\Delta K_R(\varphi)}. 
\end{equation} 
The $R_0$ values given in Table~\ref{data}, which are calculated from the $\log R/R_{\odot}(\log p,\log Z)$ calibration of \citet{marconi}, are used for determining the $R(\varphi)$ variations.}
\end{enumerate}

No error estimate of the results is given. Although it would be possible to derive  formal errors, the errors originating from the uncertainties of the synthetic light curves; from the mean magnitudes and the amplitudes  of the light curves; from the $E(I-K)$ reddening values; from applying  static atmosphere models to fit the light and colour curves of dynamic atmospheres; as well as from neglecting the $\log g$ dependence, cannot be estimated in a way that would be statistically meaningful.

\section{Results}

\subsection{Non-Blazhko RRab stars}\label{st}

The results of the process described in Section~\ref{method} are shown  for nine stable RRab stars in Fig.~\ref{lcst}.  The observed and the synthetic $I$- and $K$-band light curves  are shown in the four left-hand panels of the figure. The synthetic data are determined using an appropriately high-order ($10-15$) Fourier fit to the observations. The next two columns show the $\Delta K_{T_{\mathrm {eff}}}$ and $\Delta K_R$ curves,  disentangled from the $K$ observations as explained in items ($i$) and ($ii$) of the previous section. The estimated $R$ and $T_{\mathrm {eff}}$ variations are calculated  according to the formalism given in  item ($iii$) of Section~\ref{method}, and are plotted in the last two columns.

In order to check the reliability and correctness of the applied procedure, the derived temperature and radius changes are compared with the results of the Baade-Wesselink analysis of the M3 variables \citep{m3data} in Fig.~\ref{comp}. The amplitude ranges obtained by the process are in perfect agreement with the M3 results both for the temperature and the radius changes.  Nevertheless, the temperature  is systematically lower by $\sim300$ K for the bulge stars than the  temperature determined for the M3 stars at a given amplitude of the temperature change. This difference may arise from a $\sim0.05$ mag systematic error of the estimated $E(I-K)$ values or from the difference between the filter functions of the $K$ bands used by  the atmosphere models and by the VVV survey.

The trend seen in the amplitude  versus the mean  temperature values of the bulge sample is similar to the trend shown by the M3 stars, i.e. the hotter a star is the larger is the amplitude of its temperature change (left-hand panel of Fig.~\ref{comp}. The amplitudes of the radius change do not show such a clear pattern; the connection between $R_0$ and $A_R$ is nonlinear, and the OoI and OoII stars seem to follow different relations (right-hand panel of Fig.~\ref{comp}). However, the data of the bulge sample fit the results on the radius changes detected in M3 reasonably well, too. 

The star with the smallest-amplitude radius and temperature changes is ID\,12423. As discussed in Section~\ref{data.sec}, this star may have a significantly larger metallicy than the other stars in the sample, or its photometry may be seriously biased.

It has to be mentioned that the $K$ magnitudes are derived from aperture photometry but the $I$ magnitudes from an image-subtraction technique. Consequently, the effects of crowding, which is significant in the bulge, are not the same on the data in the two bands. The differences between the biases of crowding on the mean magnitudes and on the amplitudes in the $I$- and $K$-bands result in erroneous temperature estimates and distortions of the derived radius/temperature-change amplitudes, respectively. However, the results illustrated in Fig.~\ref{comp} indicate that the  applied method still yields reliable amplitudes, i.e. the possible uncertainties of the photometry  do not affect the results significantly.

\subsection{Blazhko stars}

\begin{figure*}
\centering
\includegraphics[width=17.7cm]{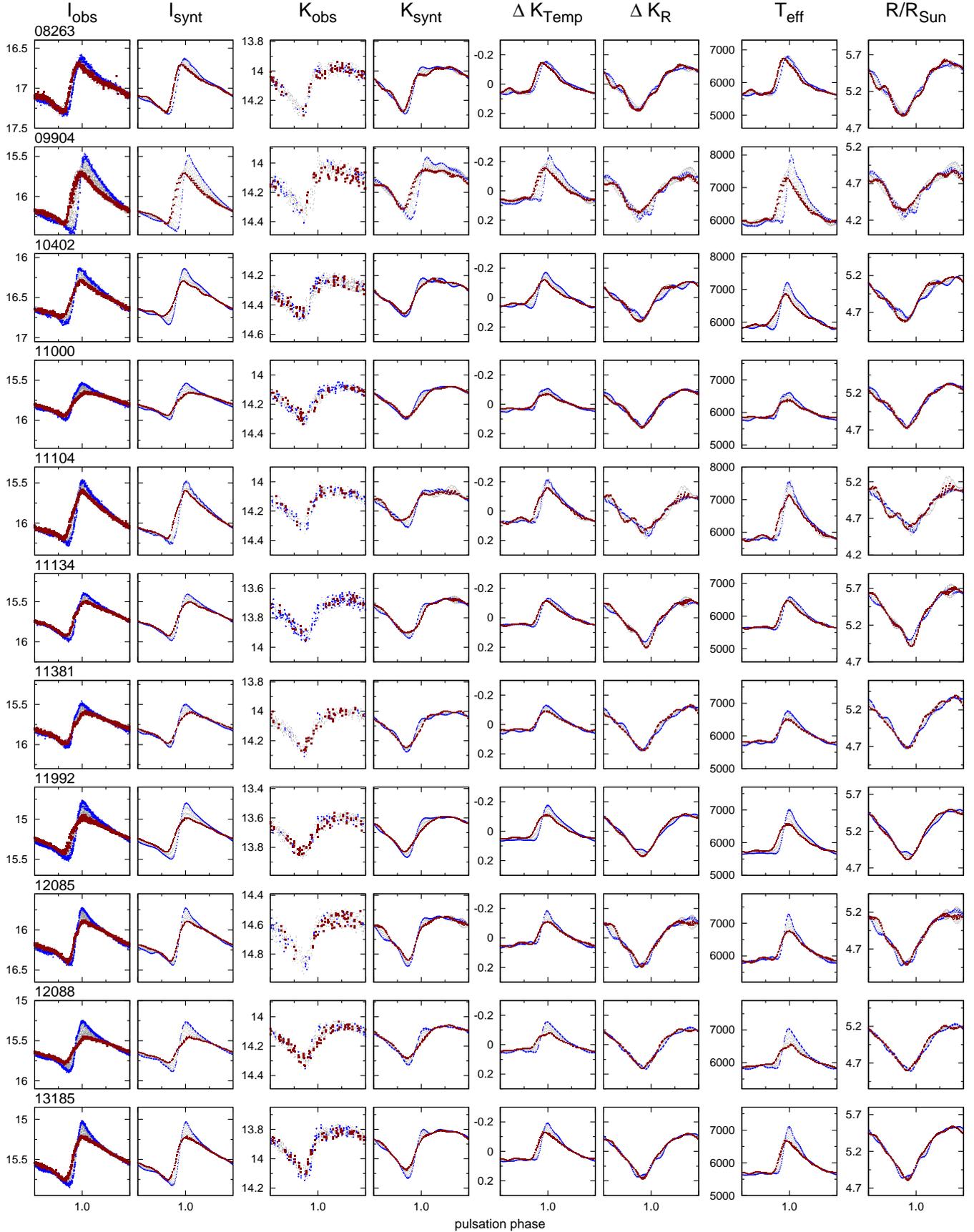}
\caption{The same plots are shown for the studied sample of Galactic-bulge Blazhko RRab stars as shown in Fig.~\ref{lcst} for the non-Blazhko reference sample. The results for the smallest- and largest-amplitude phases of the modulation are highlighted in red and blue, respectively. }
\label{lcbl} 
\end{figure*}
\begin{figure*}
\centering
\includegraphics[width=17.7cm]{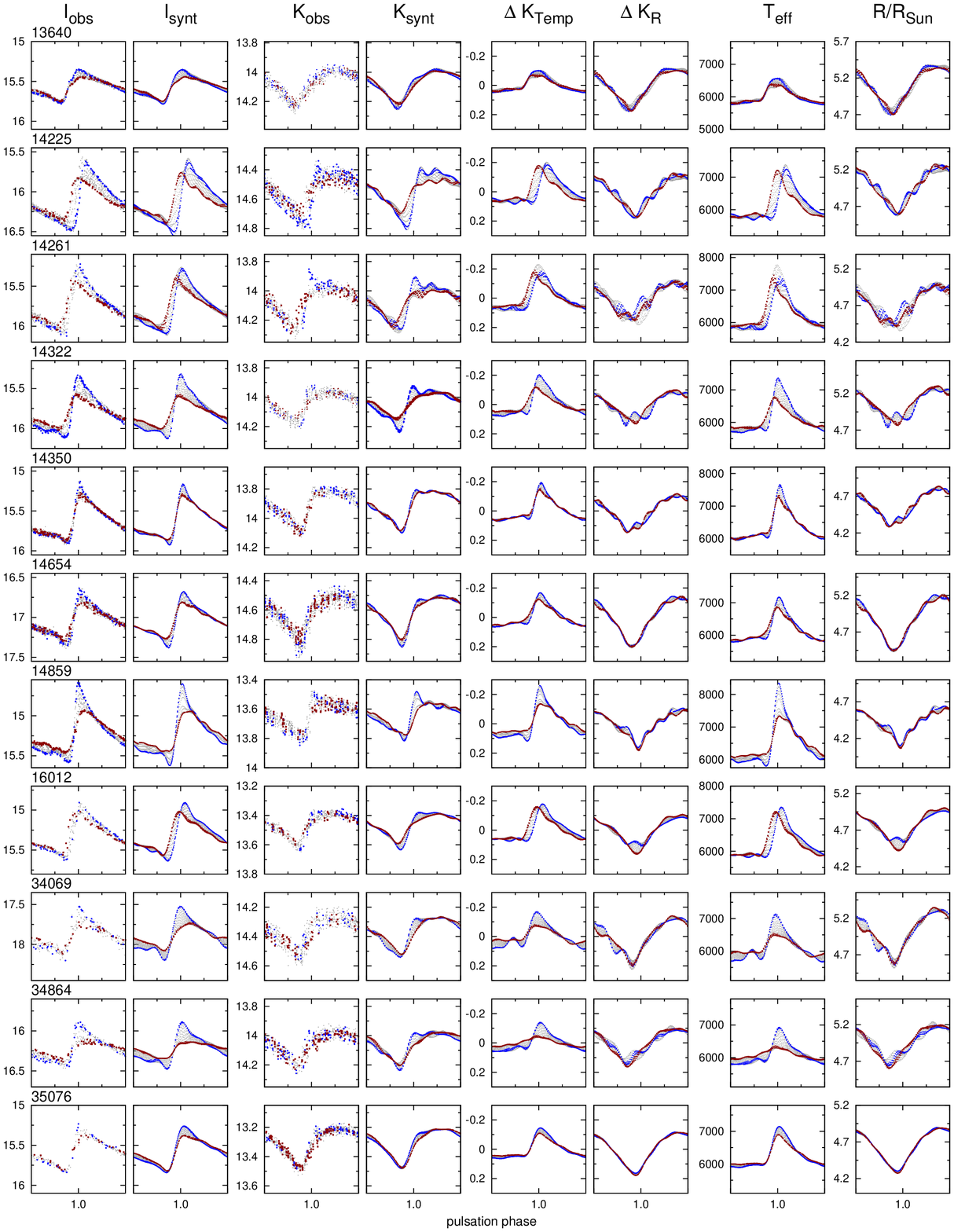}
\contcaption{}
\end{figure*}

\begin{figure*}
\centering
\includegraphics[width=17.6cm]{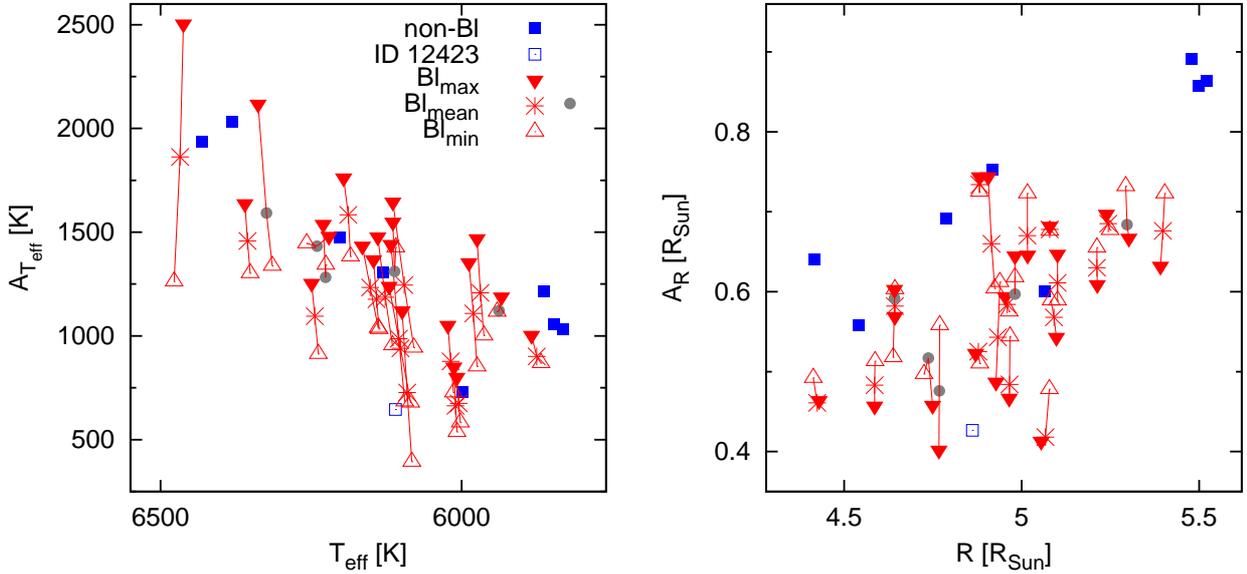}
\caption{Comparison of the full amplitudes of the temperature (left-hand panel) and radius (right-hand panel) changes of the samples of non-Blazhko and Blazhko RRab stars in the Galactic bulge. The amplitudes derived for the complete light curve and for Blazhko minimum and maximum  are indicated by stars (dots for stars with large phase modulation) and downward and upward triangles, respectively. These three points  are connected by lines for each Blazhko star.  }
\label{compbl}
\end{figure*}
\begin{figure}
\centering
\includegraphics[width=8.0cm]{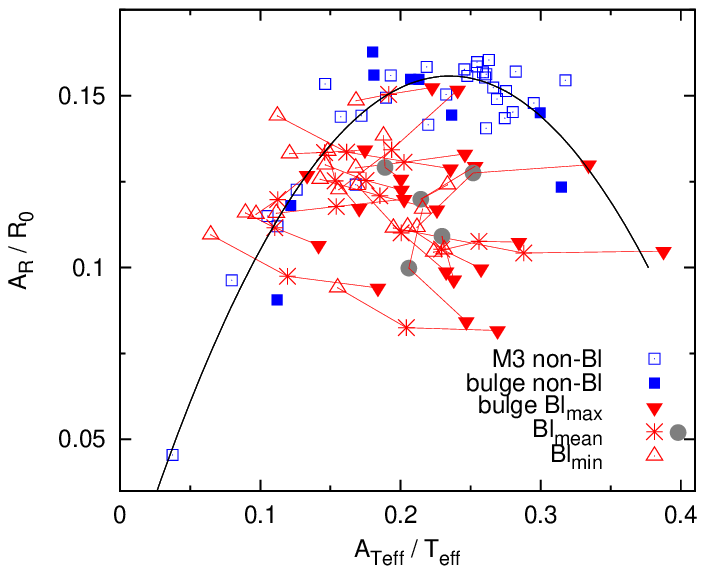}
\caption{The relation of the normalised full amplitudes of the radius and temperature changes for stable and Blazhko stars in the Galactic bulge  and for non-Blazhko stars  in the globular cluster M3  is shown. The second-order polynomial curve fitted to the non-Blazhko samples is drawn in the plot as a black curve. The mean, the minimum, and the maximum values of the amplitudes of Blazhko stars are denoted by the same symbols as in Fig.~\ref{compbl} and are connected by lines. }
\label{comprt}
\end{figure}

Blazhko stars are analysed similarly as non-Blazhko variables by using synthetic $K$ and $I-K$ data. The results for the 22 Blazhko stars studied are shown in Fig.~\ref{lcbl}. The panels are the same as shown for the non-Blazhko stars in Fig.~\ref{lcst}, the complete data sets are shown in grey colour, and data belonging to the largest- and smallest-amplitude phases of the modulation are marked in blue and red colours, respectively, in each panel of Fig.~\ref{lcbl}.

The most prominent feature of the decomposed temperature and radius variations of Blazhko stars is the lack of appreciable modulation  of the radius variations that accompany changes of the light curves.

 Neither the stars with the largest  amplitude-modulation (${{A^I}_{\mathrm {Bl_{min}}}} < 0.5 {{A^I}_{\mathrm {Bl_{max}}}}$, e.g., ID\,12088, 14859, 34069, and 34864) nor  those with the largest phase-modulation (${phase}_{\mathrm {Bl_{max}}}- {phase}_{\mathrm {Bl_{min}}} > 0.1\,P_{\mathrm {puls}}$, e.g., ID\,09904, 14225, 14261, and 16012) display notable changes in the shape of their radius-variation.  Although there might be some ambiguity in the absolute values of the determined temperature and radius variations as discussed in Section~\ref{st}, these uncertainties have no influence on this result.

The small amplitude oscillations in the radius variation seen in some cases are possibly the consequence of the uncertainty of the synthetic $K$-band light curves, caused by the improper Blazhko-phase coverage and the noise of the data. Notwithstanding, it cannot be excluded either that some minor features in the $0.9-1.2$ pulsation-phase interval, when the temperature may exceed 6750 K, arise from neglecting the $\log g$ dependence of the $K_{\mathrm{model}}$-$(I-K)$ relation as described in Section~\ref{method}. However, this phase interval corresponds to the most dynamic phase of the pulsation, when the uncertainties from using static atmosphere-model data are the largest, so any interpretation of  these features could be incorrect.

The mean  values  and the amplitudes of the radius and temperature changes are determined using the complete light curves (these are equivalent to the results obtained for the mean light curve if synthetic data are used) and also for the light-curve segments at Blazhko minimum and maximum phases for the studied 22 Blazhko stars. The mean and the full amplitudes are obtained from 8th- and 3rd-order Fourier fits of the corresponding temperature- and radius-variation curves. These data are summarised in Table~\ref{amp}. The stars exhibiting significant phase modulation are set in boldface in this table.

\begin{table*} 
\begin{center} 
\caption{Mean values and amplitudes of the radius and temperature changes of stable and Blazhko stars derived for the complete light curve and for Blazhko minimum and maximum phases.\label{amp}} 
\begin{tabular}{cc@{\hspace{5mm}}c@{\hspace{1mm}}c@{\hspace{1mm}}c@{\hspace{0mm}}c@{\hspace{5mm}}c@{\hspace{1mm}}c@{\hspace{2mm}}c@{\hspace{0mm}}c@{\hspace{5mm}}c@{\hspace{0mm}}c@{\hspace{1mm}}c@{\hspace{0mm}}c@{\hspace{3mm}}r@{\hspace{0mm}}r@{\hspace{1mm}}r}
\hline
ID$^*$&$P$ [d]&& $R_0 [R_{\odot}]$ &&&\multicolumn{3}{c}{$A_R [R_{\odot}]$}&&&$T_{\mathrm{eff}} [K]$&&&&$A_{T_{\mathrm{eff}}} [K]$&\\
\cmidrule{3-5}\cmidrule{7-9}\cmidrule{11-13}\cmidrule{15-17}
&&meanLC& Bl$_{\mathrm{min}}$ &Bl$_{\mathrm{max}}$&&meanLC& Bl$_{\mathrm{min}}$&Bl$_{\mathrm{max}}$ &&meanLC& Bl$_{\mathrm{min}}$ &Bl$_{\mathrm{max}}$ &&meanLC&Bl$_{\mathrm{min}}$&Bl$_{\mathrm{max}}$\\
\hline 
\multicolumn{17}{l}{Stable RRab stars }\\
10879& 0.470743& 4.542 &&&&0.558 &&&& 6382 &&&& 2030\\
11025& 0.684346& 5.520 &&&&0.864 &&&& 5831 &&&& 1031\\ 
11105& 0.520411& 4.787 &&&&0.691 &&&& 6203 &&&& 1474\\
12423& 0.536823& 4.862 &&&&0.426 &&&& 6110 &&&&  646\\
13714& 0.548255& 4.918 &&&&0.753 &&&& 6130 &&&& 1308\\ 
13851& 0.674377& 5.479 &&&&0.891 &&&& 5846 &&&& 1055\\ 
14408& 0.446024& 4.417 &&&&0.641 &&&& 6432 &&&& 1935\\ 
33548& 0.585278& 5.087 &&&&0.600 &&&& 5999 &&&&  731\\ 
34927& 0.679371& 5.500 &&&&0.858 &&&& 5863 &&&& 1213\\ 
\multicolumn{17}{l}{Blazhko RRab stars }\\
{\bf 08263}& 0.632747& 5.297& 5.293& 5.302&& 0.684& 0.732& 0.667&& 5938& 5941& 5934&& 1122& 1116& 1189\\
{\bf 09904}& 0.490706& 4.642& 4.638& 4.642&& 0.592& 0.518& 0.603&& 6324& 6314& 6338&& 1592& 1337& 2117\\
10402      & 0.552169& 4.933& 4.939& 4.928&& 0.543& 0.612& 0.487&& 6152& 6139& 6165&& 1234& 1039& 1433\\
11000      & 0.586359& 5.091& 5.083& 5.098&& 0.568& 0.589& 0.543&& 6010& 6008& 6013&& 663& 537& 851\\
11104      & 0.540359& 4.878& 4.882& 4.870&& 0.525& 0.510& 0.523&& 6188& 6185& 6196&& 1584& 1383& 1762\\
11134      & 0.656004& 5.398& 5.403& 5.391&& 0.676& 0.723& 0.632&& 5875& 5868& 5884&& 901& 869& 1004\\
11381      & 0.583529& 5.079& 5.079& 5.079&& 0.678& 0.676& 0.682&& 6018& 6013& 6023&& 878& 727& 1051\\
11992      & 0.613531& 5.212& 5.212& 5.213&& 0.630& 0.655& 0.609&& 5980& 5974& 5988&& 1108& 852& 1353\\
12085      & 0.547507& 4.915& 4.924& 4.906&& 0.660& 0.604& 0.744&& 6127& 6115& 6139&& 1186& 954& 1478\\
12088      & 0.557407& 4.959& 4.965& 4.954&& 0.584& 0.575& 0.594&& 6101& 6084& 6120&& 940& 678& 1240\\
13185      & 0.620459& 5.244& 5.249& 5.239&& 0.685& 0.677& 0.697&& 5969& 5963& 5974&& 1208& 1002& 1469\\
13640      & 0.588433& 5.101& 5.101& 5.101&& 0.611& 0.589& 0.647&& 6005& 6002& 6008&& 675& 582& 801\\
{\bf 14225}& 0.562246& 4.981& 4.981& 4.981&& 0.597& 0.618& 0.645&& 6121& 6119& 6124&& 1311& 1426& 1549\\
{\bf 14261}& 0.510530& 4.737& 4.725& 4.749&& 0.517& 0.497& 0.458&& 6240& 6257& 6220&& 1432& 1446& 1481\\
14322      & 0.581213& 5.067& 5.078& 5.055&& 0.418& 0.478& 0.413&& 6095& 6079& 6114&& 1245& 942& 1646\\
14350      & 0.479727& 4.586& 4.587& 4.586&& 0.483& 0.513& 0.457&& 6355& 6351& 6360&& 1458& 1302& 1638\\
14654      & 0.540359& 4.881& 4.881& 4.881&& 0.734& 0.725& 0.744&& 6141& 6137& 6146&& 1176& 1033& 1368\\
14859      & 0.447746& 4.424& 4.413& 4.429&& 0.461& 0.492& 0.464&& 6467& 6477& 6462&& 1862& 1262& 2504\\
{\bf 16012}& 0.517044& 4.768& 4.769& 4.767&& 0.476& 0.558& 0.402&& 6226& 6225& 6230&& 1282& 1343& 1538\\
34069      & 0.569616& 5.016& 5.016& 5.017&& 0.670& 0.723& 0.646&& 6104& 6094& 6117&& 986& 684& 1441\\
34864      & 0.559212& 4.966& 4.967& 4.965&& 0.484& 0.544& 0.467&& 6090& 6083& 6099&& 727& 393& 1122\\
35076      & 0.491096& 4.643& 4.643& 4.643&& 0.582& 0.603& 0.569&& 6244& 6238& 6249&& 1095& 913& 1254\\
\hline
\multicolumn{17}{l}{$^*$Blazhko stars with large-amplitude phase modulation are set in boldface. }\\
\end{tabular} 
\end{center}
\end{table*}

The  amplitudes of the derived temperature and radius changes of non-Blazhko and Blazhko RRab stars are compared in Fig.~\ref{compbl}. The full amplitude of the  variation derived for the complete light curve  and the amplitudes at Blazhko maximum and minimum are shown by different symbols. 

The  difference between the amplitudes of the temperature variation at Blazhko minimum and maximum phases is around or even larger than $1000$ K  for the stars with amplitude modulations of the largest magnitude. The amplitude of the temperature change is larger at Blazhko maximum than at Blazhko minimum for each star. High resolution spectrum analysis should be able to corroborate this result.

The amplitudes of the temperature variation ($A_{T_{\mathrm{eff}}}$) of Blazhko stars with large phase modulation, denoted by grey dots in Fig.~\ref{compbl}, show some peculiarities.
The differences between $A_{T_{\mathrm{eff}}}(\mathrm {Bl_{max}})$ and $A_{T_{\mathrm{eff}}}(\mathrm{Bl_{min}})$ are quite small for most of these stars, and the amplitude of the temperature variation derived for the mean light curve is even smaller than the $A_{T_{\mathrm{eff}}}$ amplitude detected at Blazhko minimum for some of them.   This effect indicates that using the mean light curves of Blazhko stars showing significant phase modulation  may lead to inadequate results. 

Looking at the amplitudes of the radius variations (right-hand panel in Fig.~\ref{compbl}) it seems that some  change in the amplitude of the radius variation also does occur; however, the $A_R$ amplitudes tend to be  larger at Blazhko minimum than at Blazhko maximum, i.e., this change has the opposite sign as the modulation of the light curve.  However, taking into account the uncertainties of the method (e.g., neglecting the $\log g$ dependence, using static atmosphere models, uncertainties of the $K$-band light curves, etc.) the reality of these small changes in the radius variation needs an independent verification.

Comparing the amplitudes of the temperature and radius variations of Blazhko and non-Blazhko stars   a tendency that the amplitudes of Blazhko stars are smaller than the amplitudes of non-Blazhko RRLs may be noticed in Fig.~\ref{compbl}, if the outlier, ID\,12423, is not considered. 
The $A_R/R_0$ versus $A_{T_{\mathrm {eff}}}/T_{\mathrm {eff}}$ normalised amplitude relation of stable and Blazhko stars are compared in Fig.~\ref{comprt}, in order to check whether this effect is indeed significant. For reference, the results obtained from  the Baade-Wesselink analysis of stable-light-curve RRab stars in the M3 globular cluster are also shown.
Both the bulge and the M3 non-Blazhko samples show that the amplitude of the radius change has a maximum value at around the medium value of the temperature-change amplitude.  A quadratic fit to the data of the combined sample of the bulge and the M3 non-Blazhko stars  is drawn in Fig.~\ref{comprt}.

The  amplitudes of Blazhko stars differ substantially from the ridge defined by the non-Blazhko  bulge and M3 samples, because most Blazhko stars are below this curve. The amplitudes of some of the Blazhko stars fit the positions of non-Blazhko stars the closest at Blazhko maximum, others near to Blazhko minimum or with the values determined for the mean light-curve data. Moreover, there are many variables with an anomalous relation of their  temperature- and radius-change amplitudes both at Blazhko minimum, maximum, and also as determined from the mean light curve.
Therefore, it seems that the connection between the amplitudes of the radius and temperature changes of Blazhko stars  indeed differs systematically and significantly from the relation defined by stable-light-curve stars. 

\begin{figure}
\centering
\includegraphics[width=8.0cm]{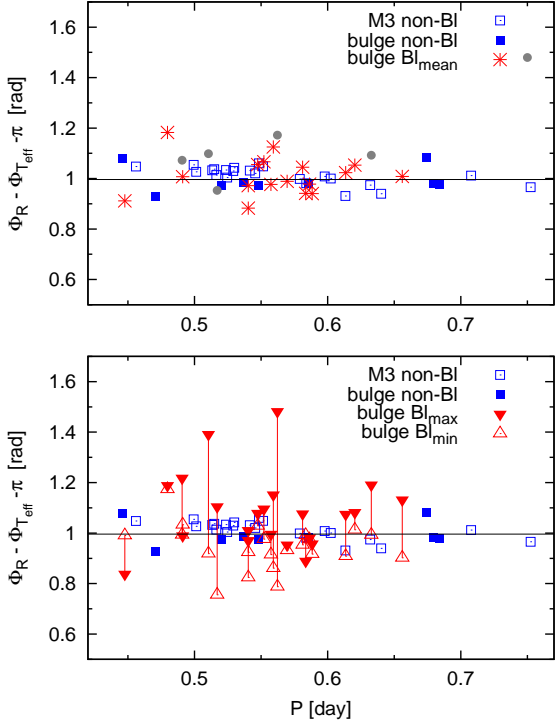}
\caption{Phase differences between the  first-order Fourier components of the radius and temperature variations. Results for the light curves of the stable-light-curve RRL stars in the bulge and in the M3 globular cluster are shown by filled and open squares, respectively. The results for the mean light curves of Blazhko stars are shown  in the top panel. The phase differences determined for Blazhko minimum- and maximum-phases (connected by vertical lines) are plotted in the bottom panel.  Grey dots denote stars with large phase modulation in the top panel. }
\label{ff}
\end{figure}

Finally, the phase relation between the derived temperature and radius changes has been determined for stable and Blazhko RRab stars, and  compared with the results obtained for the M3 non-Blazhko  stars.  Fig.~\ref{ff} shows the first-order Fourier ($sine$ decomposition) phase differences for the samples of non-Blazhko stars and for the mean light curves of Blazhko stars in the bulge (top panel) and for the minimum and maximum phases of the modulation (bottom panel). The phase differences between the radius and temperature variations of Blazhko stars cover a relatively large, $\sim0.6$ rad phase range. This is not surprising, as while the radius change of Blazhko stars remains stable, the temperature change follows the phase modulation of the light curve.
 The $\phi_R - \phi_{T_{\mathrm{eff}}}$ phase difference of Blazhko stars tends to be significantly larger and  smaller than normal at the Blazhko maximum and minimum phases, respectively. Furthermore the   phase differences determined for the mean light curves show a larger scatter around the mean value than the non-Blazhko sample.

\begin{figure}
\centering
\includegraphics[width=8.0cm]{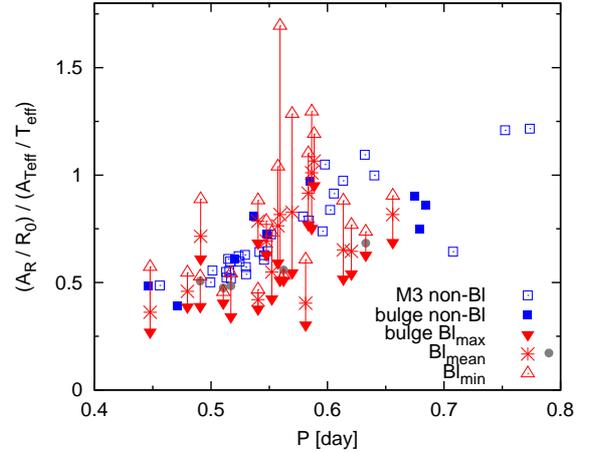}
\caption{The  relative strength of the normalised radius-change amplitude to the temperature-change  amplitude is plotted against the pulsation period for stable and Blazhko RRab stars. The full amplitudes of the variations are considered here. Data are shown for the mean, the minimum, and maximum-amplitude phases of the modulation for the Blazhko stars; the values corresponding the mean light curve of stars showing large phase modulation are denoted by grey dots. For comparison, Baade-Wesselink results of stable-light-curve RRab stars in the M3 globular cluster are also plotted. }
\label{rel.ar-at}
\end{figure}

\begin{figure}
\centering
\includegraphics[width=8.0cm]{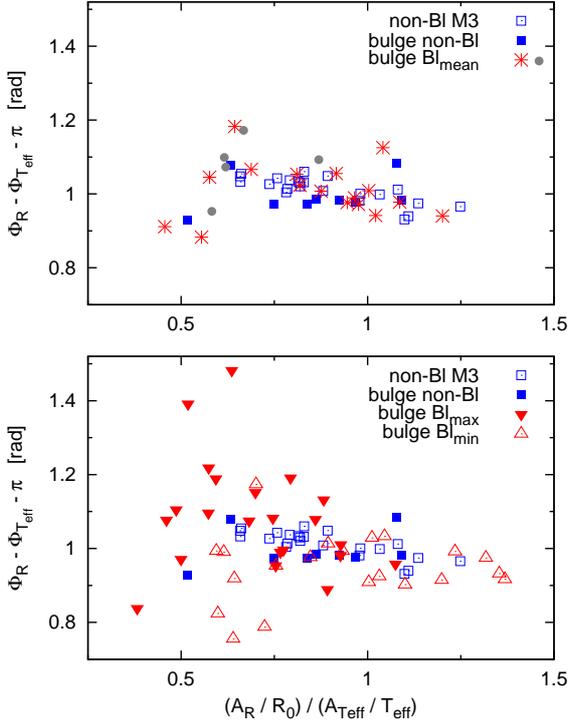}
\caption{The $\Phi_R -\Phi_{T_{\mathrm{eff}}}$ phase lag versus the relative strength of the normalised amplitude (first-order Fourier amplitude) of the radius and temperature changes is shown for the different samples of stars. The results derived from the mean light curves of Blazhko stars are compared with the data of non-Blazhko stars in the bulge and in the M3 globular cluster in the top panel.
The bottom panel shows the same sample of stars, but  data for Blazhko maximum and minimum phases are plotted here. }
\label{relmind}
\end{figure}

\section{Discussion}

The decomposition of the  temperature- and radius-change-induced contributions to the $K$-band light curve has led to the conclusion that the modulation is caused  by changes in the temperature variation and the  changes 
of the radius variation are very small and possibly non-significant. 

The lack of appreciable changes found in the radius variation parallel with the Blazhko modulation of the light curve affirms our previous result on the lack of modulation in the photometric radius variation derived from the $V$ and $I$ band light curves and Kurucz atmosphere model grids for Blazhko stars in M3 \citep{m3bw}.

It was discussed in detail in Section~\ref{amp.sec} that the relative amplitude of the  modulation is  anomalously reduced in the $K$ band.  This is the most significant in the first harmonic order. We also found that the largest discrepancy between the amplitudes of the pulsation and modulation components occurs in long-period stars (see Figs.~\ref{foura} and ~\ref{fouram}).

The lack of modulation of the radius curve helps to explain why the  $K$-band modulation amplitudes are  most reduced in the first harmonic order. According to the  decomposition of the   $K$-band light curve, it consists of a radius-change-connected sinusoidal part (that corresponds to the first-order component of a Fourier solution)  and a part connected to temperature changes, which is highly peaked (matching the higher-order components of a Fourier solution). Consequently, if the radius change is not modulated, no or only small-amplitude modulation components will be detected in the first harmonic order.

The determination of the amplitudes of the temperature and radius variations makes it possible to follow the changes of the   relative strength of these variations,  i.e. to determine whether the temperature or the radius change dominates the shaping of the observed $K$-band light curve.  

Fig.~\ref{rel.ar-at} shows the 
$(A_R/R_0) / (A_{T_{\mathrm{eff}}}/T_{\mathrm{eff}})$ ratio of the normalised relative strengths of the temperature and radius variations for the non-Blazhko samples of stars in the bulge and in M3 and also for the Blazhko variables in the bulge. Although the OoI- and OoII-type stars may show somewhat different  relations in this plot, the data of non-Blazhko stars indicate clearly  that the relative strength of the radius change  increases with increasing pulsation period, i.e. towards cooler/larger variables.  This means that the contribution of the radius change to the $K$-band light curve is increasing  and/or the  contribution of the temperature change is decreasing with increasing pulsation period. 

This finding helps to understand why the difference between the $A^K/A^I$ ratio of the pulsation and modulation frequencies is the largest for longer pulsation periods (see Fig~\ref{foura}). If the radius curve is not modulated and the contribution of the radius change to the light curve is larger at long periods than at short periods, than it  follows naturally that a larger discrepancy in the amplitudes of  the modulation frequencies  will be detected at longer periods.

The location  of the mean parameters of Blazhko stars matches the region covered by non-Blazhko stars in Fig.~\ref{rel.ar-at} reasonably well. The relative strength of the radius change to the temperature change is larger at Blazhko minimum than at Blazhko maximum for each of the studied Blazhko stars. This is a direct consequence of the stability of the amplitude of the radius change, parallel with the large changes detected in the amplitude of the temperature variation. 
The relative strength of the radius change to the temperature change seems to be anomalous for most of the Blazhko stars both at Blazhko minimum and at Blazhko maximum.

 The non-adiabaticity of the excitation mechanism of the pulsation and also the dynamics of the partial ionisation zones modify the phase-lags between the direct observables in classical radial mode pulsators \citep[]{c68,c71,szabo07}.
Likewise, the variations of photospheric radius and temperature with time in RRLs, and the phase-lags between them, are affected by non-adiabaticity.

Fig.~\ref{ff} indicates that the $\Phi_R-\Phi_{T_{\mathrm{ eff}}}-\pi$ phase-lag is around 1.0, which corresponds to $\sim0.15$ pulsation phase.
In the adiabatic case, $\Phi_R-\Phi_{T_{\mathrm{ eff}}}$ should be equal to $\pi$, because the radius and temperature are showing opposite behaviour. 
(Note here that the phase-lags between the first-order Fourier components are different from those determined between the extrema.)

However, for all Blazhko stars showing large-amplitude phase modulation, the phase lags are smaller at Blazhko minimum than at maximum, as can be seen in Fig~\ref{ff}. Hence, our results hint that non-adiabaticity has a smaller effect at Blazhko minimum than at Blazhko maximum, which would be a natural consequence if the excitation of the pulsation was weaker when the amplitude of pulsation is smaller.

The question arises then, whether there is any connection between the   phase-lag values  and the  relative strengths of the radius change compared to temperature change, and also whether it depends on the Blazhko phase. 
Therefore, finally we inspect how the $\Phi_R -\Phi_{T_{\mathrm{eff}}}$ phase lags are related to the relative  strength of the normalised amplitude of the radius and temperature changes in Blazhko stars. The results for the mean light curves and for the Blazhko minimum and maximum phases are compared to the data of non-Blazhko stars in the top and bottom panels of  Fig~\ref{relmind}, respectively. 

The distribution of the M3 and bulge samples of stable RRab stars does not show any significant differences. 
The positions of Blazhko stars corresponding to the mean light curve show a bit larger scatter than of the stable stars, but without any systematic trend. The radius change of the Blazhko stars at the maximum phase of the modulation tends to have smaller relative strengths compared to the strength of the temperature change than it has in stable stars or at Blazhko minimum. On the other hand, systematic differences in the phase lags seem to be dominant at Blazhko minimum, as the phase lags tend to be smaller than normal in this modulation phase.

\section{Summary}

The analysis of the VVV $K$-band light curves combined with the OGLE-IV $I$-band  observations  has led to the detection of Blazhko modulation in the $K$  band. However, the strength of the modulation is anomalously reduced in the $K$ band. 

In order to explain the features of the modulation in the $K$ band (see details in Section~\ref{amp.sec}), as well as  to interpret the results in the context of the lack of modulation in the photometric radius curves of Blazhko stars  \citep{m3bw}, a method has been developed to decompose the $K$-band light curves according to the physical contributions from the radius  and the temperature change of the photosphere.

Based on the results of this method, the following conclusions on the modulation are drawn, and should be taken into account in any interpretation of the Blazhko phenomenon.
\begin{itemize}
\item{The radius variation of the photospheric regions associated with the pulsation  does not show significant changes in Blazhko stars. The analysis of both the Galactic bulge $I$- and $K$-band data (this paper) and the M3 globular cluster $V$- and $I$-band data \citep{m3bw} have led to this conclusion.}
\item{Modulation of the temperature change curve is responsible for the observed modulation of the light curve.} 
\item{The stability of the radius variation of the photosphere demonstrates the stability of the pulsation. It excludes any interpretation of the phenomenon that connects the detected phase changes of the light curve with changes in the period of the pulsation.  }
\item{Both the phase and the amplitude relations between the radius and temperature changes of the photosphere are anomalous in most phases of the modulation for most of the Blazhko stars. The ($\Phi_R -\Phi_{T_{\mathrm{eff}}}$) phase lags tend to be smaller than normal at Blazhko minimum while the relative strength of the radius change to the temperature change during the pulsation cycle tends to be smaller than normal at Blazhko maximum.  }
\item{Based on the phase lags between the radius and temperature changes it seems  that non-adiabatic effects are more important in large-amplitude phases of the modulation than when the amplitude is small.}
\end{itemize}

From all these painstakingly derived clues it appears that we need to look for a mechanism to explain the Blazhko effect that periodically/cyclically modifies the  phase (phase-lag) and amplitude connection between the  radius and temperature changes of the photosphere during the pulsation cycle, while  the period of the pulsation remains stable.

\section*{Acknowledgments}
G.H. acknowledges support from the Graduate Student Exchange Fellowship Program between the Institute of Astrophysics of the Pontif\'icia Universidad Cat\'olica de Chile and the {\it Zentrum f\"ur Astronomie} of the University of Heidelberg, funded by the Heidelberg Center in Santiago de Chile and the {\it Deutscher Akademischer Austauschdienst (DAAD)},  and by CONICYT-PCHA/Doctorado Nacional grant 2014-63140099. I.D. and E.K.G. were supported by Sonderforschungsbereich SFB 881 ``The Milky Way System'' (subproject A3) of the German Research Foundation (DFG). Additional support for G.H. and M.C. is provided by the Ministry for the Economy, Development, and 
Tourism's Millennium Science Initiative through grant IC\,120009, awarded to the Millennium 
Institute of Astrophysics (MAS); by Proyecto Basal PFB-06/2007; by FONDECYT grant \#1171273; and 
by CONICYT's PCI program through grant DPI20140066.

\bibliographystyle{mnras}

\label{lastpage}
\end{document}